\documentclass[a4paper,11pt]{article}
\usepackage{pos}
\usepackage{graphicx}
\usepackage{wrapfig}
\usepackage{ascmac}
\usepackage{bm}
\usepackage{amsmath}
\usepackage{tikz}
\usepackage{tikz-feynman}
\usetikzlibrary{decorations.pathmorphing}
\usetikzlibrary{arrows.meta}
\usetikzlibrary{bending}
\tikzset{snake it/.style={decorate, decoration=snake}}
\tikzset{gluon/.style={decorate, decoration={snake, amplitude=1pt, segment length=4pt}, thick}}

\title{Complex potential and open system applications in heavy-ions and cold atoms}

\author*[a]{Yukinao Akamatsu}

\affiliation[a]{Department of Physics, Osaka University,\\
Toyonaka, Osaka 560-0043, Japan}

\emailAdd{yukinao.a.phys@gmail.com}

\abstract{
Since the discovery of the complex potential of quarkonium at high temperatures, quarkonium has been regarded as an open quantum system in the quark-gluon plasma. 
Recently, a similar issue regarding in-medium bound states of impurities has also emerged in particle physics and cold atomic physics. 
We will provide an overview of recent advancements in understanding key quantities such as complex potential and transport coefficients for heavy impurities in finite temperature QCD and cold atomic systems.
}

\FullConference{The XVIth Quark Confinement and the Hadron Spectrum Conference (QCHSC24)\\
 19-24 August, 2024\\
 Cairns Convention Centre, Cairns, Queensland, Australia\\}

\begin{document}
\maketitle

\section{Introduction}
There have been various advancements achieved in the study of heavy-ion collisions. 
Among the various milestones, we have successfully created a deconfined state of matter known as the quark-gluon plasma (QGP) and uncovered its unexpectedly strong coupling nature, characterized by a remarkably low shear viscosity to entropy density ratio ($\eta/s$). 
Nevertheless, several significant questions remain unresolved. 
For instance, the mechanisms underlying the rapid thermalization of the system during its swift expansion are still not fully understood. 
Additionally, the phenomenon of hydrodynamic collectivity observed in systems with a relatively small number of particles poses an ongoing challenge to our understanding. 
Finally, our knowledge of the dynamical properties of strongly coupled QGP remains limited.

Quarkonium, a bound state of a heavy quark and its antiquark, serves as an ideal probe to investigate the color dynamics in the QGP. 
As a localized object with color charge, quarkonium is expected to reflect the modifications of the color force in the medium. 
In the deconfined phase, the potential between quarks becomes screened and short-ranged due to the liberation of color degrees of freedom, leading to the dissociation of bound states at sufficiently high temperatures. 
This phenomenon forms the basis of the $J/\psi$ suppression scenario~\cite{Matsui:1986dk}, which suggests that the production of $J/\psi$ particles decreases when QGP is created in heavy-ion collisions.

\begin{figure}
\centering
\includegraphics[width=0.4\textwidth]{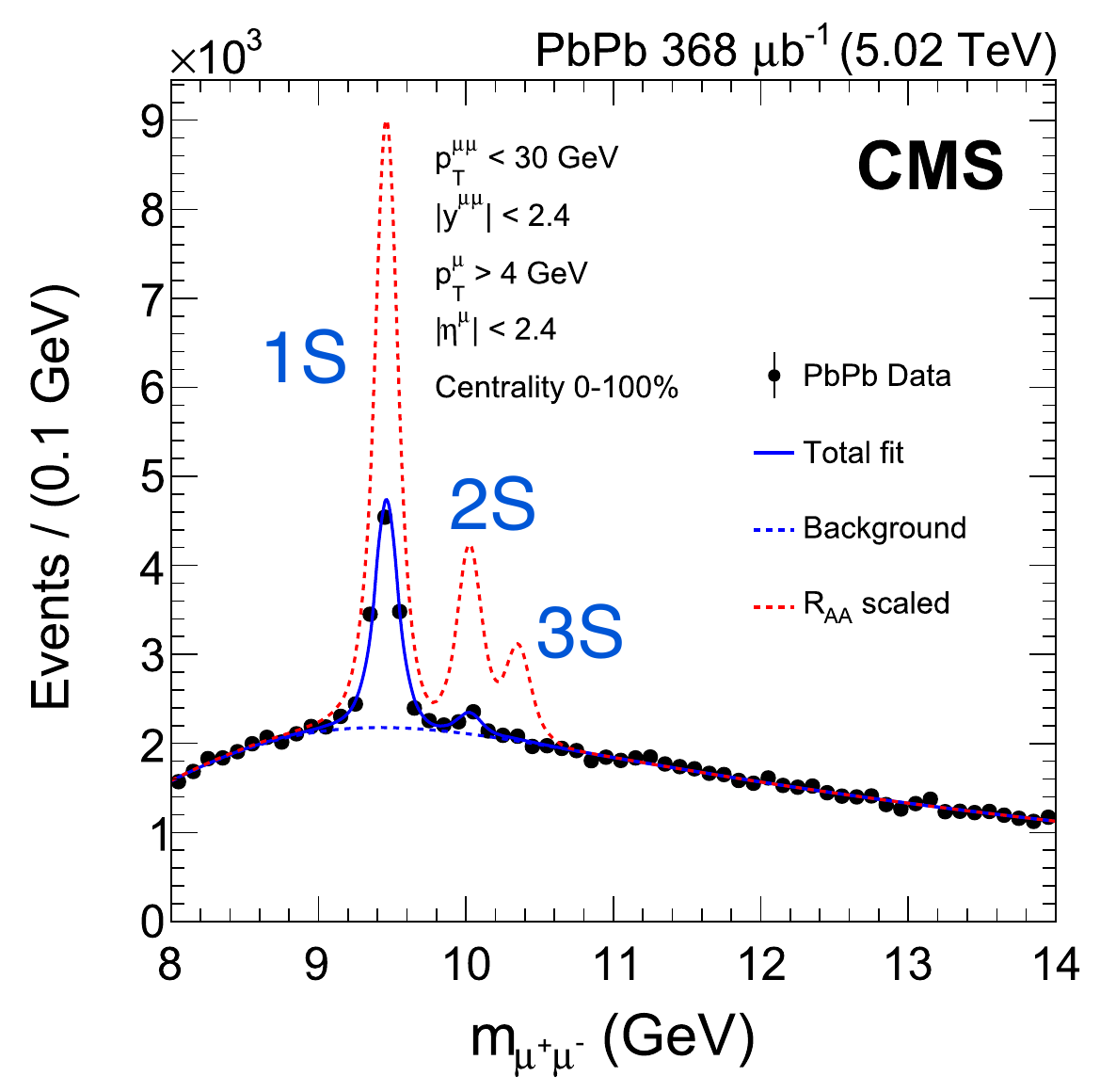}
\caption{Dimuon invariant mass spectrum around $\Upsilon$ masses from the CMS collaboration at the LHC.
Data show clear pattern of sequential suppression, where higher excited states are more strongly suppressed.
Figure adapted from Ref.~\cite{CMS:2018zza}.}
\label{fig:cms}
\end{figure}

Consider, for example, the experimental data of dimuons containing $\Upsilon$ peaks observed at the LHC~\cite{CMS:2018zza}, which compares heavy-ion collision data with proton-proton (pp) collision data in Fig.~\ref{fig:cms}. 
The findings indicate that excited states are suppressed more strongly than the ground state. 
This sequential melting can be qualitatively explained using the screening model. 
However, dynamical effects, such as collisions and gluon absorptions or emissions, also play a significant role. 
Recent theoretical studies suggest that, within the potential framework, these dynamical effects manifest in the imaginary part of the potential~\cite{Laine:2006ns}, while static screening effects are represented in the real part. 
Since the imaginary part of the potential should not lead to a reduction in the heavy quark number, it is crucial to establish a theoretical framework that integrates both static and dynamical effects, enabling a more comprehensive understanding of the QGP.

\section{Complex potential for quarkonium}
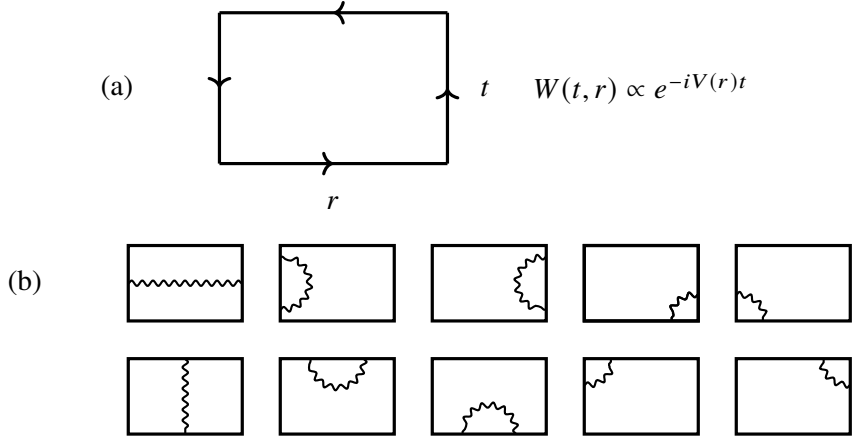
\begin{figure}
\begin{center}
\begin{tikzpicture}
    \node (end) at (-1,1) [left] {(a)};
    \draw[very thick, ->] (0,0) -- (1.5,0);
    \draw[very thick, -] (1.4,0) -- (3,0);
    \draw[very thick, ->] (3,0) -- (3,1);
    \draw[very thick, -] (3,0.9) -- (3,2);
    \draw[very thick, ->] (3,2) -- (1.5,2);
    \draw[very thick, -] (1.6,2) -- (0,2);
    \draw[very thick, ->] (0,2) -- (0,1);
    \draw[very thick, -] (0,1.1) -- (0,0);
    \node (end) at (1.5,-0.3) [below] {$r$};
    \node (end) at (3.3,1) [right] {$t$};
    \node (end) at (4,1) [right] {$W(t,r)\propto e^{-iV(r)t}$};
\end{tikzpicture}\\
\vspace{3mm}
\begin{tikzpicture}
    \node (end) at (-1,0.5) [left] {(b)};
    \draw[very thick] (0,0) rectangle (1.5,1);
    \draw[very thick, gluon] (0,0.5) -- (1.5,0.5);
    \draw[very thick] (2,0) rectangle (3.5,1);
    \draw[very thick, gluon] (2,0.15) arc (-90:90:0.35);
    \draw[very thick] (4,0) rectangle (5.5,1);
    \draw[very thick, gluon] (5.5,0.85) arc (90:270:0.35);
    \draw[very thick] (6,0) rectangle (7.5,1);
    \draw[very thick, gluon] (7.5,0.35) arc (90:180:0.35);  
    \draw[very thick] (6,0) rectangle (7.5,1);
    \draw[very thick, gluon] (7.5,0.35) arc (90:180:0.35);  
    \draw[very thick] (8,0) rectangle (9.5,1);
    \draw[very thick, gluon] (8,0.35) arc (90:0:0.35);  
    \draw[very thick] (0,-1.5) rectangle (1.5,-0.5);
    \draw[very thick, gluon] (0.75,-0.5) -- (0.75,-1.5);
    \draw[very thick] (2,-1.5) rectangle (3.5,-0.5);
    \draw[very thick, gluon] (3.1,-0.5) arc  (0:-180:0.35);
    \draw[very thick] (4,-1.5) rectangle (5.5,-0.5);
    \draw[very thick, gluon] (5.1,-1.5) arc  (0:180:0.35);
    \draw[very thick] (6,-1.5) rectangle (7.5,-0.5);
    \draw[very thick, gluon] (6,-0.85) arc  (-90:0:0.35);
    \draw[very thick] (8,-1.5) rectangle (9.5,-0.5);
    \draw[very thick, gluon] (9.5,-0.85) arc  (-90:-180:0.35);
\end{tikzpicture}
\end{center}
\caption{(a) Real-time Wilson loop $W(t,r)$ at finite temperature. Its long time behavior determines the complex potential between static heavy quark pair.
(b) Leading order perturbative expansion of $W(t,r)$ in the soft regime $r\sim 1/gT$. The gluon propagators are dressed with the Hard-Thermal-Loop self energies.}
\label{fig:WilsonLoop}
\end{figure}

We review the key insights gained about the complex potential.
Its definition is given in terms of a static heavy quark pair. 
\begin{align}
\langle\Psi(\bm r, t)\rangle_T = \langle Q_c(\bm  0, t)Q(\bm r, t)Q^{\dagger}(\bm r, 0)Q_c^{\dagger}(\bm 0, 0)\rangle_T
\xrightarrow[t\to\infty]{} e^{-iV(\bm r)t}.
\end{align}
Here $Q_{(c)}$ annihilates a heavy quark (antiquark) and $Q_{(c)}^{\dagger}$ creates a heavy quark (antiquark).
The pair wave function $\Psi(\bm r, t)$ contains the information of potential energy in its phase. 
At the same time, due to the thermal fluctuations, this potential energy possesses a stochastic nature. 
Taking the medium average $\langle\Psi(\bm r, t)\rangle_T$, phase cancellation occurs and gives rise to the imaginary part of the potential $V(\bm r)$. 
By integrating out the static quarks in the singlet state (by connecting between $Q$ and $Q_c$ with a Wilson line), what remains is the real-time thermal Wilson loop, whose long-time behavior defines the potential (Fig.~\ref{fig:WilsonLoop}).

\subsection{Complex potential in perturbation theory}
In perturbation theory, the potential in the soft regime ($r\sim 1/gT$) can be calculated using Hard-Thermal-Loop resummed propagators. 
The resulting potential contains both real and imaginary parts~\cite{Laine:2006ns, Beraudo:2007ky, Brambilla:2008cx}:
\begin{align}
\label{eq:potential_LO}
V_{\rm 1LO}(r)=  -\frac{C_F g^2}{4\pi}\left(m_D + \frac{e^{-m_{\rm D}r}}{r}\right)
-i C_F g^2 T\int\frac{d^3k}{(2\pi)^3}\frac{\pi m_{\rm D}^2(1-e^{i\bm k\cdot\bm r})}{k(k^2+ m_{\rm D}^2)^2},
\end{align}
with the Debye mass $m_D = gT\sqrt{\frac{N_c}{3}+\frac{N_f}{6}}$ and $C_F=(N_c^2-1)/2N_c$.
The real part shows mass shift and screening, while the imaginary part arises from the Landau damping due to the collisions.
At $1/T \ll r\ll 1/m_D$, the binding energy is $\sim g^2/r$ while the decay rate is $\sim g^2 T m_D^2 r^2$.
Therefore the resonance peaks exist only when $r\lesssim 1/g^{2/3}T$.

\begin{figure}
\centering
\includegraphics[width=0.7\linewidth]{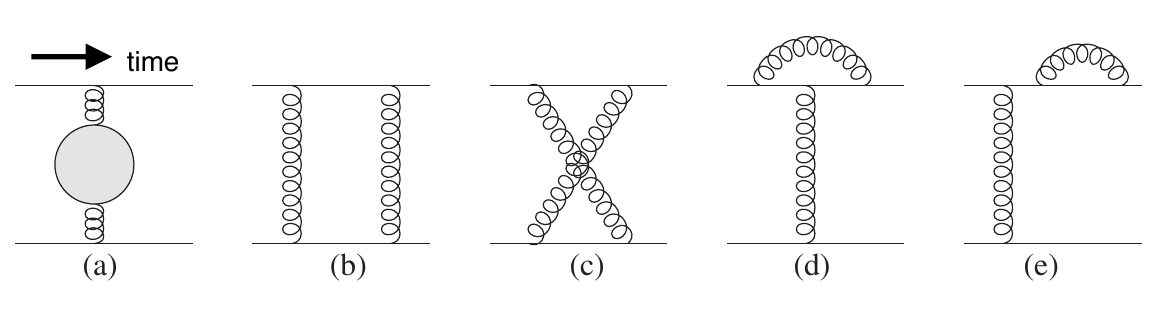}
\caption{
Diagrams for the complex potential in the next-to-leading order expansion.
Figure adapted from Ref.~\cite{Carrington:2024ize}.
}
\label{fig:potential_NLO}
\end{figure}

Recently, next-to-leading order (NLO) calculation has been performed in a semi-hard regime $r\sim 1/g^a T$ with $1/3 < a < 2/3$~\cite{Carrington:2024ize}.
The NLO contribution stems from one-loop diagrams ((b)-(e) in Fig.~\ref{fig:potential_NLO}) and corrections to one-loop self-energy beyond the HTL approximation ((a) in Fig.~\ref{fig:potential_NLO}). 
In the $p$-space, the NLO contribution reads
\begin{align}
\Delta \tilde V_{\rm 2, exp}(p) = - \frac{g^4 C_F N_c T}{16\pi m_D p^2} 
\left\{\begin{aligned}
& \left[ 1 - \frac{3\pi^2}{16} + \frac{4\pi m_D}{p} 
+ \frac{m_D^2}{p^2} \left( \frac{5\pi^2}{24} - \frac{4}{3} \right) \right] \\
& + i\frac{\pi T m_{D}}{p^2} \left[ \frac{56}{3\pi}  
- \left( 1 - \frac{3\pi^2}{16} \right) \frac{m_D}{p} 
- \left( 1 - \frac{N_f}{2N_c} \right) \frac{4p}{\pi T}\right] 
\end{aligned}\right\},
\end{align}
where expansion with respect to $m_D/p \ll 1$ is performed.
The NLO improved complex potential can be used calculate the self-energy of bottomonia, which is compared with corresponding lattice QCD simulations.

\subsection{Complex potential by the lattice QCD simulations}
The potential has also been calculated by lattice QCD simulations. 
A key challenge arises because the Euclidean formalism does not provide direct access to real-time data. 
However, the imaginary-time Wilson loop and the real-time Wilson loop are related through a common spectral function via analytic continuation, from the Fourier transform to the Laplace transform:
\begin{align}
W(t=-i\tau,r) = \int d\omega e^{-\omega \tau}\rho(\omega,r), \quad 0\leq \tau \leq \beta .
\end{align}
In lattice QCD simulations, a finite set of data points along imaginary time $W(t=-i\tau_n, r) \ (n=1,2, \cdots, N_{\tau})$ is obtained, including statistical errors.
The spectral function is then inferred using the Bayesian reconstruction method. 
Once the spectral function $\rho(\omega, r)$ is reconstructed, $W(t,r)$ can be obtained, from which the potential $V(r)$ is extracted.
A study conducted in 2015~\cite{Burnier:2014ssa} found that the real part of the potential exhibits screening, while the imaginary part increases monotonically with $r$.
However, contrary to natural and conventional expectations, the latest study~\cite{Bazavov:2023dci} reports no screening in the real part of the potential.
In this recent study, instead of using the Bayesian method, the lattice data $W(t=-i\tau_n, r)$ is fitted using an ansatz spectral function:
\begin{align}
\rho(\omega, r) = \rho_{\rm low}(\omega, r) + \rho_{\rm peak}(\omega, r) + \rho_{\rm high}(\omega, r).
\end{align}
Here, $\rho_{\rm high}(\omega, r)$ is a temperature-independent component that dominates at large $\omega$.
It is fixed by a spectral function at $T=0$ with $\rho_{T=0}(\omega, r) = A(r)\delta(\omega - V_{T=0}(r)) + \rho_{\rm high}(\omega, r)$.
The term $\rho_{\rm peak}(\omega, r)$ represents a temperature-dependent peak with a finite width, which determines the complex potential $V(r)$ at finite temperature.
Finally, $\rho_{\rm low}(\omega, r)$ corresponds to contributions well below the peak and is interpreted as a heavy $Q\bar Q$ state propagating forward in Euclidean time while interacting with a backward propagating light state from the medium.
These two studies present conflicting results regarding the nature of screening in the deconfined phase of QCD matter. 
In particular, this discrepancy raises the question of how screening should manifest in a nonperturbative and non-static framework, where peak broadening occurs and the distinction between different peaks (such as $\rho_{\rm low}$ and $\rho_{\rm peak}$) becomes ambiguous.

\section{Quarkonium as an open quantum system in QGP}
The presence of an imaginary component in the potential indicates that quarkonium is not a closed system but rather an open system in the quark-gluon plasma (QGP). 
We now introduce the open-system approach to quarkonium in the QGP~\cite{Akamatsu:2020ypb}.

Open quantum systems are ubiquitous. 
Theoretically, we begin with a total closed system, whose Hilbert space is a direct product of the system and its environment. 
Our primary interest lies in system observables, which can be computed using the reduced density matrix $\rho_S$.
The master equation governing the reduced density matrix follows the Lindblad form when we impose positivity ($\rho_S>0$) and trace conservation (${\rm Tr}\rho_S=1$) of the density matrix:
\begin{align}
\frac{d\rho_S}{dt} &= -i[H_S', \rho_S] + \sum_{k}L_k\rho_S L_k^{\dagger} - \frac{1}{2}L_k^{\dagger}L_k\rho_S
- \frac{1}{2}\rho_S L_k^{\dagger}L_k \\
&= -iH_{\rm eff}\rho_S + i\rho_S H_{\rm eff}^{\dagger} + \sum_{k}L_k\rho_S L_k^{\dagger},
\quad H_{\rm eff} = H_S' -\frac{i}{2}\sum_k L_k^{\dagger}L_k .
\label{eq:non-Hermitian+jump}
\end{align}
The Lindblad equation can also be expressed differently as in \eqref{eq:non-Hermitian+jump}, incorporating a non-Hermitian Hamiltonian $H_{\rm eff}$ and quantum jumps $\sum_{k}L_k\rho_S L_k^{\dagger}$. 
The complex potential can be interpreted as a component of the non-Hermitian Hamiltonian.
Thus, for a complete description, it is essential to account for the quantum jumps dictated by the Lindblad operators $L_k$.

A few remarks are in order. 
The Lindblad equation assumes initially uncorrelated states for the total system $\rho_{\rm tot}(0) = \rho_S\otimes\rho_E$, which may be a reasonable assumption for heavy quark pair production in hard partonic processes. 
Additionally, its microscopic derivation presumes weak system-environment coupling. 
However, this assumption may not hold for quarkonium in a strongly coupled QGP. 
Nevertheless, studying quarkonium dynamics in an idealized scenario provides valuable insights.

\subsection{Lindblad equation for $r\sim gT$ and $g\ll 1$}
When quarkonium and the QGP interact perturbatively, the Lindblad equation can be derived directly from the non-relativistic quantum mechanical Hamiltonian of heavy quarks ($\bm x$) and antiquarks ($\bm x_c$):
\begin{align}
H_I &= g\left[A^a_0(\bm x) (t^a\otimes 1) - A^a_0(\bm x_c) (1\otimes t^{a*})\right] \\
&=\int\frac{d^3k}{(2\pi)^3}\left[e^{i\bm k\cdot\bm x} (t^a\otimes 1)- e^{i\bm k\cdot\bm x_c}(1\otimes t^{a*})\right]
\otimes g\tilde A^a_0(\bm k),
\label{eq:qm_NRQCD}
\end{align}
which can also be expressed in Fourier space as in \eqref{eq:qm_NRQCD}. 
For instance, at a distance of $1/gT$, the leading-order Lindblad operator in the recoilless limit takes a form similar to the Hamiltonian~\cite{Akamatsu:2014qsa}:
\begin{align}
&L_{\bm k}^a = \sqrt{\gamma_k}\Bigl[e^{i\bm k\cdot\bm x}(t^a\otimes 1) -  e^{i\bm k\cdot\bm x_c}(1\otimes t^{a*})\Bigr] 
+ \mathcal O(\dot {\bm x}, \dot {\bm x_c}) 
\end{align}
Here, ``recoil" refers to changes in the heavy quark states during collisions, implying an implicit assumption that heavy quarks move slowly. 
In the Lindblad operators, $k$ and $a$ serve as labels, while $\bm x$ and $t^a$ are operators. 
The Lindblad operator describes scattering events with momentum transfer $\bm k$ and color rotation in channel $a$. 
The environmental effects are averaged out through the two-point function, which determines the Lindblad operator's coefficient $\gamma_k$:
\begin{align}
\gamma_k\delta^{ab}(2\pi)^3\delta(\bm k-\bm k') 
&= g^2\int dt\langle \tilde A_0^a(\bm k,t)\tilde A_0^b(-\bm k',0)\rangle_{T, k\sim gT} \nonumber \\
&=  \frac{\pi g^2 T m_{\rm D}^2}{k(k^2+ m_{\rm D}^2)^2}\delta^{ab}(2\pi)^3\delta(\bm k-\bm k').
\end{align}
Notably, we assumed specific conditions for $g\ll 1$ and $r\sim 1/gT$, allowing the Lindblad operators to reproduce the imaginary part of the (singlet) potential \eqref{eq:potential_LO} by $-\frac{1}{2}\sum_{a}\int \frac{d^3k}{(2\pi)^3} L_k^{a\dagger}L_k^a$.

\subsection{Lindblad equation for quarkonium dipole}
When the quarkonium size is small, the dipole approximation can be used without assuming a small coupling constant 
$g$. 
In this limit, an effective field theory known as potential NRQCD (pNRQCD) is applicable, where the non-relativistic interaction Hamiltonian takes the following form:
\begin{align}
H_I &=- r_i\Bigl[
\sqrt{\frac{1}{2N_c}}\left(|a\rangle\langle s| + |s\rangle\langle a|\right)
+ \frac{1}{2}d^{abc}|b\rangle\langle c| 
\Bigr] \otimes g E_i^a(\bm R) 
+ (T_A^a)_{bc}|b\rangle\langle c|\otimes gA^a_0(\bm R).
\end{align}
The last term can be removed by field redefinition in the path integral formulation~\cite{Brambilla:2016wgg, Brambilla:2017zei}.
In addition to the standard electric dipole interaction, color dynamics also play a role. 
The interaction consists of singlet-octet transitions and octet-octet processes, where $d$ represents a symmetric structure constant of the SU(3) color group. 
In the leading order of $r$, the Lindblad operator can be inferred from this interaction Hamiltonian~\cite{Brambilla:2016wgg, Brambilla:2017zei}: 
\begin{align}
L_i^a &= \sqrt{\gamma}r_i\Bigl[
\sqrt{\frac{1}{2N_c}}\left(|a\rangle\langle s| + |s\rangle\langle a|\right) + \frac{1}{2}d^{abc}|b\rangle\langle c|
\Bigr] + \mathcal O(\dot r).
\end{align}
Since the details of the octet space are not our primary concern, we trace it out to obtain independent Lindblad operators for each transition: singlet-to-octet, octet-to-singlet, and octet-octet processes:
\begin{align}
L_i^{os} &= \sqrt{\frac{\gamma_{os}}{2N_c}}r_i |o\rangle\langle s| , \quad
L_i^{so} = \sqrt{\frac{\gamma_{so}}{2N_c}}r_i |s\rangle\langle o| , \quad
L_i^{oo} = \sqrt{\frac{\gamma_{oo}}{4}}r_i |o\rangle\langle o| . 
\end{align}
Because this derivation depends on the smallness of the dipole size $r$ rather than the coupling strength $g$, the coefficients of the Lindblad operators are defined non-perturbatively.

From the previous example, it is evident that the transport coefficients are defined in terms of the two-point functions of the color electric fields. 
However, these two-point functions are not gauge-invariant. 
To address this issue, one must carefully analyze field redefinitions, which we will not elaborate on here. 
In essence, the field redefinition amounts to shifting the octet basis back to the infinite past. 
To express the result in the original local basis, adjoint Wilson lines must be inserted. 
For instance, in the singlet-to-octet transition rate, the electric fields are connected by an adjoint Wilson line $U_{A}(t_1,t_2) $ representing the octet quarkonium~\cite{Scheihing-Hitschfeld:2023tuz}:
\begin{align}
&\gamma_{os} = \frac{g^2}{3(N_c^2-1)}\int dt\langle E_i^a(t)U^{ab}_{A}(t,0) E_i^b(0)\rangle_{T} 
= (N_c^2-1)\gamma_{so}, \\
& U_{A}(t_1,t_2) = {\rm P}\exp\left[
-i\int_{t_1}^{t_2} dt gA^a(\bm R, t)(T_A^a)
\right].
\end{align}
This differs from the heavy quark momentum diffusion constant, where fundamental Wilson lines appear between the electric fields~\cite{Casalderrey-Solana:2006fio, Caron-Huot:2008dyw}:
\begin{align}
&\kappa = \frac{g^2}{3N_c}\int dt\langle {\rm Tr}U_{F}(-\infty, t)E_i(t)U_{F}(t,0)E_i(0)U_{F}(0,-\infty)\rangle_{T}, \\
& U_{F}(t_1,t_2) = {\rm P}\exp\left[
-i\int_{t_1}^{t_2} dt gA^a(\bm R, t)(T_F^a)
\right].
\end{align}
The octet-octet transition rate is more closely related to the heavy quark momentum diffusion constant $\kappa$.
The primary distinction is that the electric fields are connected via adjoint Wilson lines:
\begin{align}
&\gamma_{oo} = \frac{g^2}{3(N_c^2-1)}\int dt
\langle {\rm Tr}U_{A} (-\infty,t) \mathcal E_i(t)U_{A} (t,0)\mathcal E_i(0)U_{A} (0,-\infty)\rangle_T, \\
& d^{abc}E^a_i =:(\mathcal{E}_i)_{bc}.
\end{align}
This can be understood by comparing the interaction Hamiltonians of octet quarkonium and single heavy quarks:
\begin{align}
H_I^{oo} &=- \frac{1}{2} \bm r \cdot g \bm E^a(\bm R) d^{abc} |b\rangle\langle c|,\\
H_I^{Q} &= -\bm r\cdot g\bm E^a(\bm x)(t^a)_{ij}|i\rangle\langle j|.
\end{align}
Consequently, two transport coefficients ($\gamma_{os}$ and $\gamma_{oo}$) characterize the Lindblad equation in the dipole limit, rather than just one, as was initially assumed in early open-system studies.

\subsection{Thermalization of quarkonia}
\begin{figure}
\centering
\includegraphics[width=0.4\linewidth]{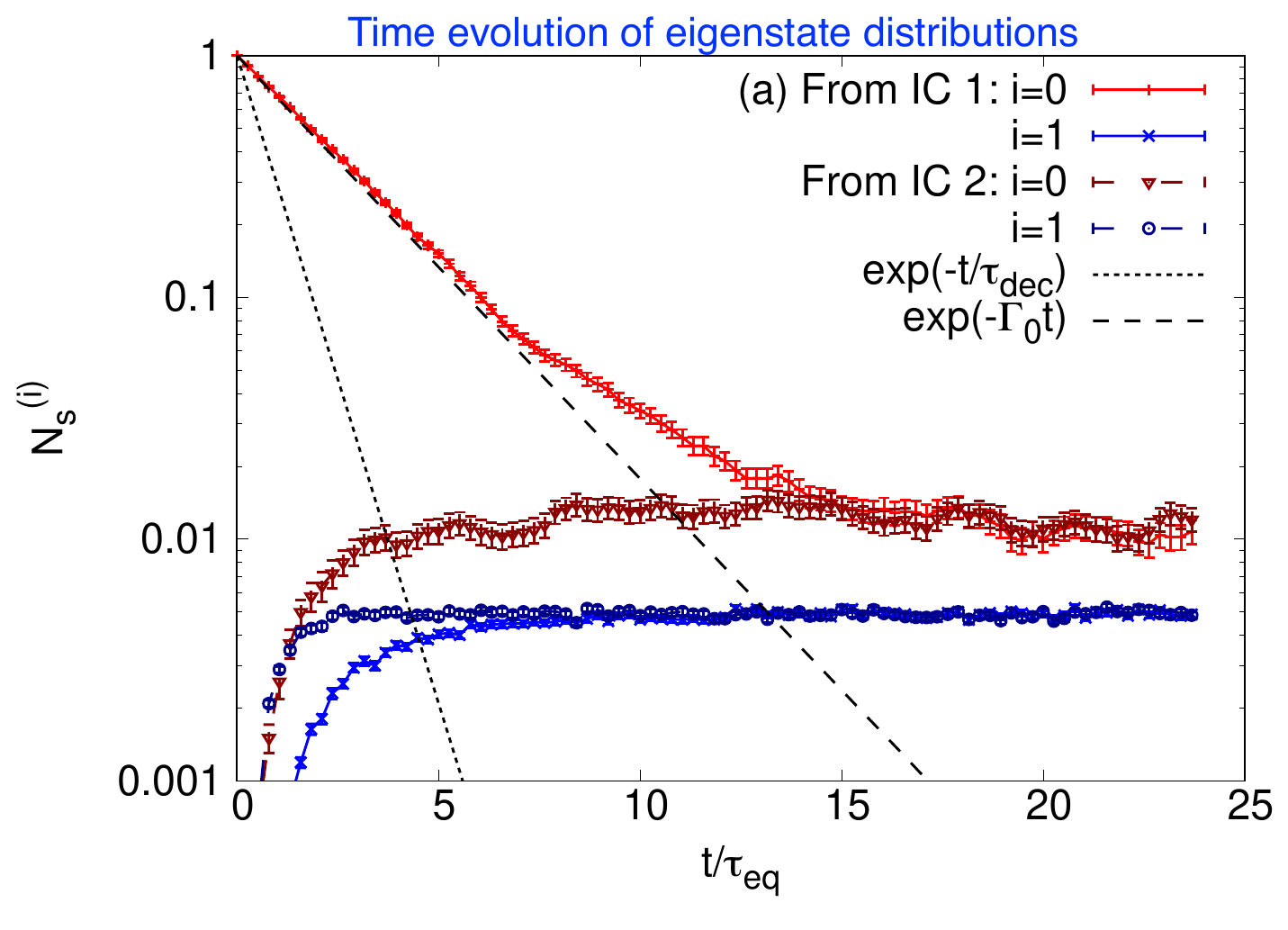}
\includegraphics[width=0.4\linewidth]{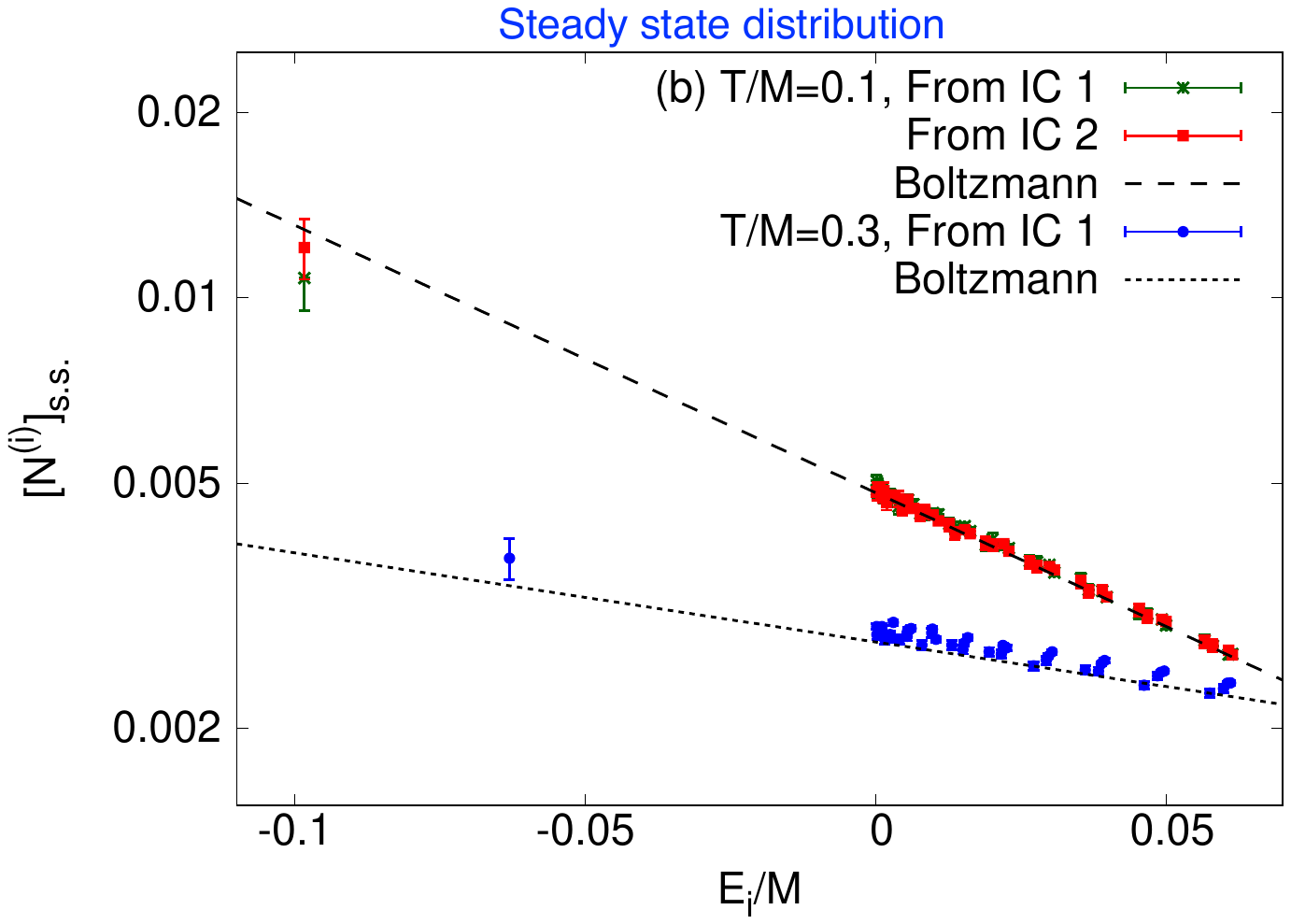}
\caption{
(a) Time evolution of the eigenmode distribution for the lowest two levels (i=0,1).
For both initial conditions (IC1 and IC2), the distribution reaches a steady state.
(b) Eigenmode distributions at the steady state.
They are consistent with the Boltzmann distribution at the medium temperature ($T=0.1M$ and $0.3M$).
Figures adapted from Ref.~\cite{Miura:2022arv}.
}
\label{fig:equilibration}
\end{figure}

The dynamics of quarkonium in the QGP is a combination of quantum Brownian motion and simultaneous color transitions. 
In heavy-ion collisions, this process is frequently simulated for $\Upsilon$ mesons.
Let us discuss our numerical results from a one-dimensional simulation at fixed temperatures~\cite{Miura:2019ssi, Miura:2022arv}. 
We solved the Lindblad equation using a stochastic unraveling method, in which mixed-state wave functions are sampled:
\begin{align}
\rho_S(x,y,t) = \lim_{N\to\infty}\frac{1}{N}\sum_{i=1}^N \psi_i(x,t)\psi_i^*(y,t).
\end{align}
For further details, we refer to the original paper. 
Our simulations considered two distinct initial conditions.
\begin{itemize}
\item Singlet Ground State Initialization (IC1): 
This scenario corresponds to a short formation time for quarkonium. 
Initially, the singlet ground state undergoes dipole-induced excitations to the octet state.
Over time, comparable to the relaxation time, the density matrix approaches a diagonal form and reaches a steady state.
\item Octet Wave Packet Initialization (IC2): 
This scenario corresponds to a long formation time or early thermalization of the QGP. 
Here, the singlet density matrix also exhibits a structure due to dipole transitions. 
Again, over time, the density matrix diagonalizes and reaches a steady state.
\end{itemize}
To assess quarkonium thermalization, we computed the eigenstate occupation numbers (Fig.~\ref{fig:equilibration}). 
Regardless of the initial conditions, the distribution converges to a steady state. 
At equilibrium, the distribution aligns with the Boltzmann distribution at the environmental temperature.

Thermalization occurs when first-order recoil effects are included. 
A schematic explanation is as follows.
Consider a simple interaction Hamiltonian $H_I = V_S\otimes V_E$.
From this, we derive the corresponding Lindblad operator, as described earlier:
\begin{align}
L = \sqrt{\gamma}\left( V_S + \frac{i}{4T} \dot V_S + \cdots \right) \propto V_S - \frac{1}{4T}[H_S, V_S] +\cdots .\end{align}
The recoil effect is incorporated via a derivative expansion in time. 
Using this Lindblad operator with first-order recoil, we calculate transition amplitudes between system eigenstates $|\epsilon_i\rangle$:
\begin{align}
\langle \epsilon_2|L|\epsilon_1\rangle 
&\propto \langle \epsilon_2|V_S|\epsilon_1\rangle \left(1-\frac{\epsilon_2-\epsilon_1}{4T}\right), \\
\frac{\Gamma_{1\to2}}{\Gamma_{2\to1}} &= \frac{|\langle \epsilon_2|L|\epsilon_1\rangle|^2}{|\langle \epsilon_1|L|\epsilon_2\rangle|^2}
=\left(\frac{1-\frac{\epsilon_2-\epsilon_1}{4T}}{1-\frac{\epsilon_1-\epsilon_2}{4T}}\right)^2
\simeq \exp\left(-\frac{\epsilon_2-\epsilon_1}{T}\right), \\
& \because 
\quad \left(\frac{1+x/4}{1-x/4}\right)^2 \simeq 1+x+\frac{1}{2}x^2+\frac{3}{16}x^3 + \cdots \simeq e^x.
\end{align}
The ratio of forward and backward rates then satisfies an approximate detailed balance condition. Without recoil, the forward and backward rates remain identical, preventing proper thermalization and leading to excessive heating instead.

\section{Complex potential for polarons in cold atoms}
Finally, we explore a novel application of the open quantum system framework to polarons in cold atomic gases.
The concept of a polaron was first introduced by Landau and Pekar~\cite{Landau:1948ijj}. 
In metals, a conduction electron polarizes the surrounding crystal lattice, forming a quasiparticle that consists of both the electron and the phonons it excites. 
This quasiparticle is known as a polaron, an effect typically neglected in band theory. 
In general, the effective mass of a polaron can differ significantly from that of a free electron.

In the cold atom community, a polaron broadly refers to an impurity particle immersed in a quantum gas. 
By selecting different atomic species, researchers can fine-tune mass ratios and adjust interactions between the impurity and the surrounding gas particles, allowing the polaron to exhibit either attractive or repulsive behavior. 
Due to the high degree of experimental controllability, polarons in cold atomic gases provide an ideal platform for simulating quarkonium physics.

\subsection{Scaling of the imaginary potential at long distance}
Our recent work investigates the impurity potential in the superfluid phase of a cold atomic gas. 
In this regime, low-energy excitations—phonons ($\varphi$)—can be described using effective field theory~\cite{Son:2005rv}:
\begin{align}
    \label{eq:EFT}
    \mathcal L_{\rm eff} &= p(\theta) + \Phi^{\dagger}\left(i\partial_t + \frac{1}{2M}\bm\nabla^2\right)\Phi
    -gn(\theta)\Phi^{\dagger}\Phi, \\
    \theta &= \mu - \partial_t\bar\varphi - \frac{1}{2m}(\bm\nabla\bar\varphi)^2,
\end{align}
where we assume a contact interaction $\propto n(\theta) \Phi^{\dagger}\Phi$ between the impurity ($\Phi$) and the gas particles.
Expanding the pressure $p(\mu)$ and the number density $n(\mu)=p'(\mu)$ at $T=0$ with respect to the phonon field $\varphi \equiv \sqrt{\chi}\bar\varphi$, where $\chi(\mu) = n'(\mu)$, the effective Lagrangian reads
\begin{align}
    \mathcal L_{\rm eff} &= \mathcal L_{\rm ph}(\varphi) + \mathcal L_{\rm pol}(\Phi) 
    + g\left[\sqrt{\chi}\partial_t\varphi + \frac{1}{2m}(\bm\nabla\varphi)^2\right]\Phi^{\dagger}\Phi,\\
    \mathcal L_{\rm ph} &= p(\theta) = \frac{1}{2}(\partial_t\varphi)^2 - \frac{1}{2}c_s^2(\bm\nabla\varphi)^2
    +\frac{1}{2m\sqrt{\chi}}(\partial_t\varphi) (\bm\nabla\varphi)^2 
    +\frac{1}{8m^2\chi}((\bm\nabla\varphi)^2)^2 +\cdots,\\
    \mathcal L_{\rm pol} &= \Phi^{\dagger}\left(i\partial_t + \frac{\bm\nabla^2}{2M} - g n\right)\Phi.
\end{align}

\begin{figure}
\centering
\includegraphics[width=0.25\linewidth]{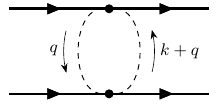}
\caption{
Two-phonon exchange diagram.
}
\label{fig:two-phonon}
\end{figure}

We calculate the induced potential between polarons and find that, since this potential arises from zero-energy transfer processes, one-phonon exchange does not contribute due to the presence of time derivatives. 
Instead, the leading contribution comes from two-phonon exchange (Fig.~\ref{fig:two-phonon}), yielding the imaginary part of the potential as
\begin{align}
 V_{\rm Im}(\bm r) &= 
 - \frac{2\pi g^2}{m^2} \int \frac{d^3k}{(2\pi)^3}\frac{d^3q}{(2\pi)^3} 
 e^{i(\bm k-\bm q) \cdot \bm r} 
 \frac{(\bm q\cdot\bm k)^2}{4 E_{\bm k}^2}
 \delta ( E_{\bm k}- E_{\bm q})  
 [1+n_B(E_{\bm k})] n_B (E_{\bm k}) ,
\end{align}
where $E_{\bm k} = c_s|\bm k|$ is the on-shell energy of phonon.
Note that from here, the imaginary part of the potential means $V_{\rm Im}(r) - V_{\rm Im}(\infty)$ in terms of the previous notation.
Performing the angular integration with $g \simeq 2\pi a_{\rm{IM}}/m$, we obtain the following result:
\begin{align}
 V_{\rm Im} (r) 
 &=  - \frac{a_{\rm{IM}}^2 T^7}{2\pi m^4 c_s^{10}} h (k_T r),
\end{align}
where we introduced a thermal phonon momentum scale $k_T \equiv T/c_s$ and the function $h (y)$
\begin{align}
 h (y) &\equiv 
 \int_0^\infty ds\,
 \frac{ s^6 e^{s}}{(e^s-1)^2}  \left[
  \frac{3j_1 (sy)^2 }{(sy)^2} 
  - \frac{2j_1 (sy) j_2 (sy)}{sy} 
  + j_2 (sy)^2
 \right],
\end{align}
with the spherical Bessel functions $j_n(x)$.
At long distances and finite temperatures, the real part of the potential scales as  $1/r^6$ due to the massless nature of phonons~\cite{Fujii:2022oun}, while the imaginary part scales as $1/r^2$~\cite{Akamatsu:2023yus}.

\begin{figure}
\centering
\includegraphics[width=0.3\linewidth]{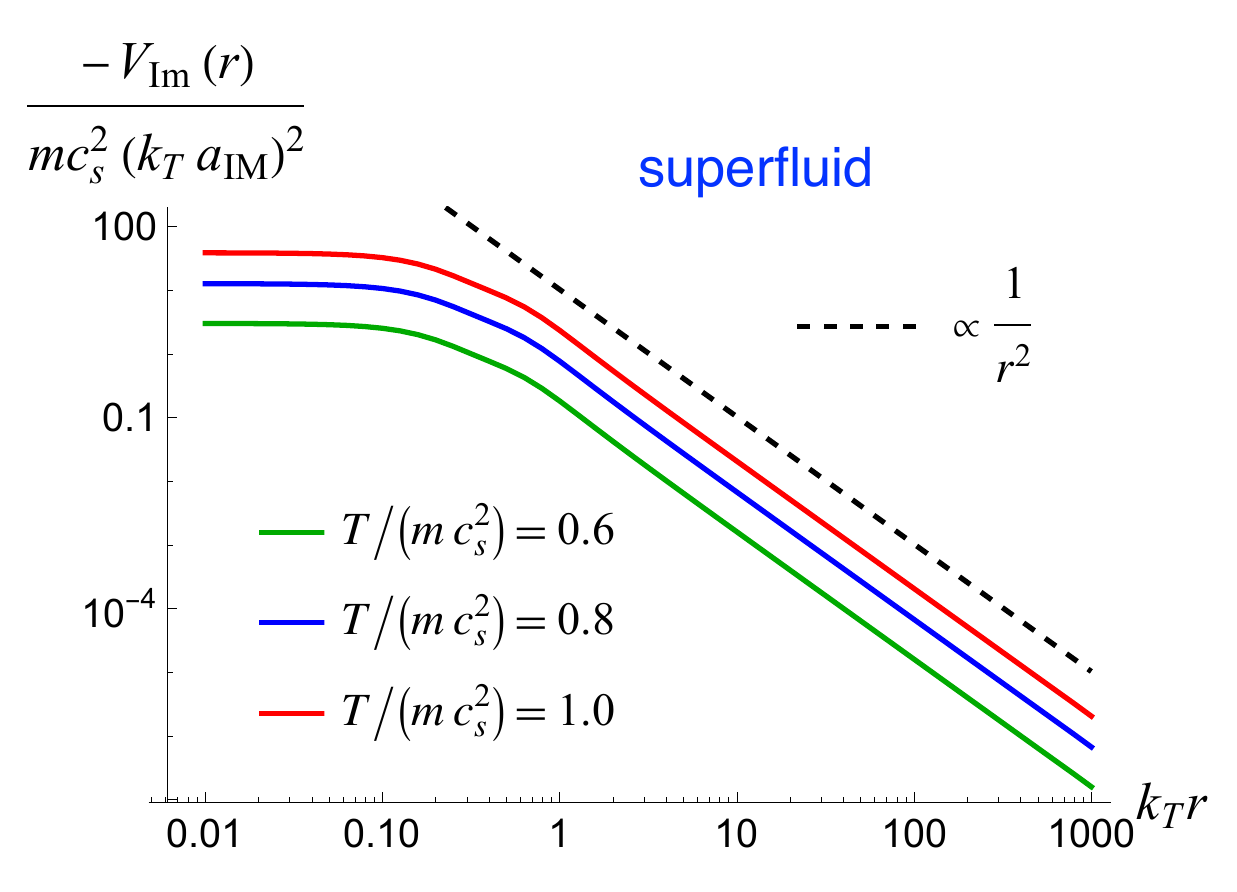}
\includegraphics[width=0.3\linewidth]{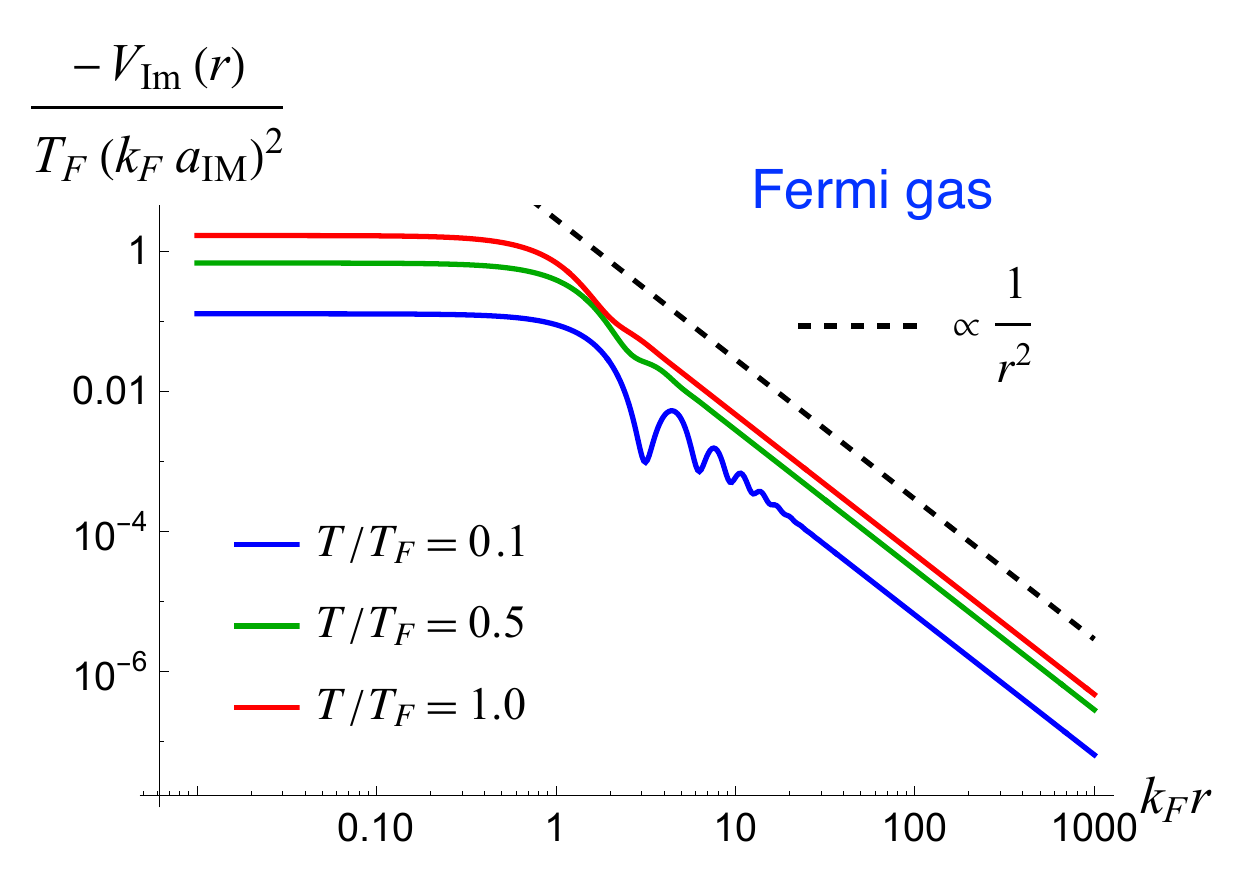}
\includegraphics[width=0.3\linewidth]{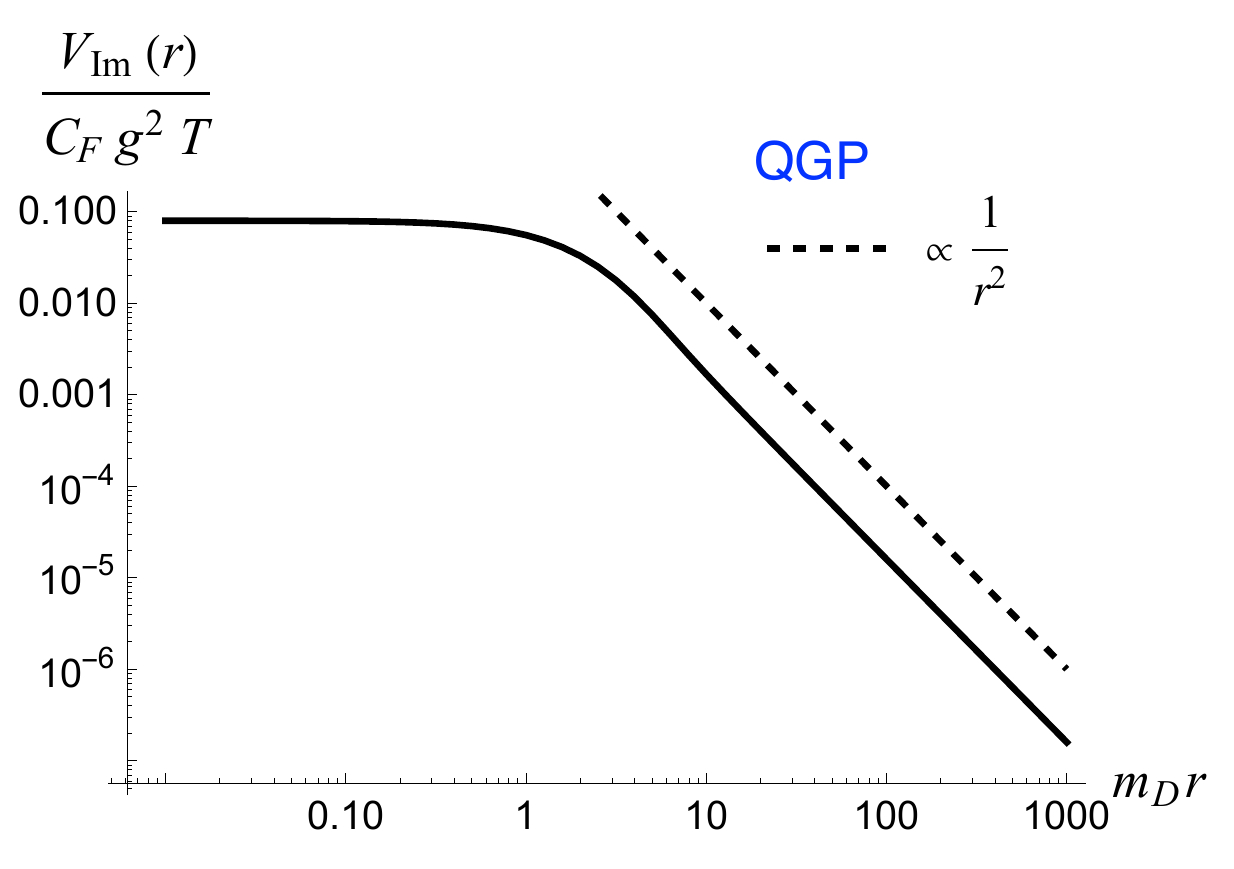}
\caption{
Scaling of the imaginary part of the potential $\propto 1/r^2$ at long distances.
The systems considered are superfluid (left), Fermi gas (middle), and QGP (right).
Figures adapted from Ref.~\cite{Akamatsu:2023yus}.
}
\label{fig:scaling}
\end{figure}

A key question arises: What is the underlying physical mechanism behind the scaling behavior of the imaginary part?
Our numerical results for the imaginary potential in a superfluid reveal the same $1/r^2$ scaling (Fig.~\ref{fig:scaling}). 
It is tempting to attribute this behavior to the presence of gapless excitations. 
Interestingly, a similar scaling is observed in a free Fermi gas, where gapless particle-hole excitations are present.
However, a crucial counterexample challenges this interpretation: quarkonium in the QGP. Despite the absence of massless excitations in the QGP, the imaginary potential still exhibits the same  $1/r^2$ scaling.
This suggests that the observed behavior arises from a more general mechanism, independent of the presence of gapless modes.

\subsection{Scaling in the collisional regime}
This universal behavior can be explained by a simple physical mechanism. 
In all these systems, the imaginary potential originates primarily from two-body scattering processes:
\begin{align}
\widetilde{V}_{\rm Im} (\bm k) &\propto  
 - \int_{\bm q}  |\mathcal M_{\bm k+\bm q, \bm q}|^2 \delta (E_{\bm k+\bm q} - E_{\bm q})
n (E_{\bm q}) \big[ 1 \pm n (E_{\bm k+\bm q}) \big],
\end{align}
where $E_{\bm q} = c_s |\bm q|$ for superfluid phonons, $E_{\bm q} = \frac{\bm q^2}{2m}$ for free fermi gas, and $E_{\bm q} = |\bm q|$ for QGP.
To obtain an instantaneous potential, energy conservation must be imposed in the collision kinematics. 
This introduces a delta function constraint $\delta (E_{\bm k+\bm q} - E_{\bm q})\to \delta(\cos \theta_{\bm k\bm q})/v_{\bm q} k \ (k\to 0)$, which contributes a $1/k$ factor at small momentum $k$, while other factors remain approximately constant. 
Consequently, the imaginary potential scales as $1/k$ in momentum space, leading to a universal $1/r^2$ dependence in real space within the collisional regime.
This fundamental insight provides a unifying explanation for the observed universality of the imaginary potential across diverse physical systems.

\section{Summary}
We explore the application of the open quantum system framework to quarkonium dynamics in the quark-gluon plasma (QGP) and extend the discussion to polarons in cold atomic gases.
We begin by introducing the concept of open quantum systems, where quarkonium interacts with the QGP as an environment. 
The Lindblad equation is used to describe this interaction, incorporating a non-Hermitian Hamiltonian and quantum jumps. The complex potential emerges naturally from this framework. 
We discuss different approximations, such as the weak coupling limit and dipole limit, which allow a systematic treatment of quarkonium interactions. 
The transport coefficients in the Lindblad equation for the latter are related to gauge-invariant correlation functions, and their structure is analyzed with an emphasis on the octet sector.
Next, we present numerical simulations of quarkonium evolution using stochastic unraveling of the Lindblad equation. 
We consider different initial conditions—starting from either a singlet ground state or an octet wave packet—and analyze the thermalization process. 
The eigenstate occupation number confirms that the system reaches a steady-state distribution consistent with a Boltzmann distribution at the environmental temperature. 
The importance of recoil effects in achieving thermalization is also demonstrated.
Finally, we extend the discussion to polarons in cold atomic gases, highlighting their relevance as analog systems for studying quarkonium physics. 
Using effective field theory, we calculate the impurity potential in a superfluid and show that at long distances, the imaginary part scales universally as $1/r^2$. 
We explore the origin of this scaling, initially linking it to the presence of gapless excitations. 
However, by comparing with quarkonium in the QGP—where no massless excitations exist yet the same scaling appears—we identify a deeper, more universal mechanism rooted in two-body scattering processes.
This study provides new insights into the dynamics of open quantum systems in high-energy and condensed matter physics, bridging the gap between quarkonium evolution in the QGP and impurity dynamics in ultracold atomic gases.

\section*{Acknowledgements}
The author is supported by Japan Society for the Promotion of Science (JSPS) KAKENHI Grant Number JP23K25870.
He is also deeply indebted to the late Prof. Tetsuo Matsui, who pioneered the research field of $J/\psi$ suppression and provided us with warm support.

\end{document}